\documentclass[
showpacs,
floatfix,
aps,
prb,
amsmath,
twocolumn,
superscriptaddress, 
tightenlines
]{revtex4}

\usepackage{graphicx}
\usepackage{dcolumn}
\usepackage{amsmath}
\usepackage{amsfonts}
\usepackage{bm}
\usepackage{epsfig}
\usepackage{times}
\usepackage[varg]{txfonts}
 
\newcommand{\be}{\begin{equation}}
\newcommand{\ee}{\end{equation}}
\newcommand{\bea}{\begin{eqnarray}}
\newcommand{\eea}{\end{eqnarray}}

\newcommand{\po}{{\mathcal P_\omega}}

\def\xt{\mathcal{X}_{2}}
\def\eac{\epsilon_{\mathrm{ac}}}

\def\oc{\omega_{\mbox{\scriptsize {c}}}}

\def\tauq{\tau_{\mbox{\scriptsize {q}}}}

\def\vsp{\vspace{-0.06 in}}

\newcommand{\req}[1]{Eq.~(\ref{#1})}

\newcommand{\oncite}[1]{Ref.~\onlinecite{#1}}

\bibpunct{}{}{,}{s}{,}{,}
\begin{document}

\title{
Giant microwave photoresistivity in high-mobility quantum Hall systems
}

\author{A.\,T. Hatke}
\affiliation{School of Physics and Astronomy, University of Minnesota, Minneapolis, Minnesota 55455, USA}
\author{M.\,A. Zudov}
\email[Corresponding author: ]{zudov@physics.umn.edu}
\affiliation{School of Physics and Astronomy, University of Minnesota, Minneapolis, Minnesota 55455, USA}
\author{L.\,N. Pfeiffer}
\affiliation{Department of Electrical Engineering, Princeton University, Princeton, NJ 08544, USA}
\author{K.\,W. West}
\affiliation{Department of Electrical Engineering, Princeton University, Princeton, NJ 08544, USA}

\received{28 October 2010; revised manuscript received 4 February 2011; published 7 March 2011}

\begin{abstract}
We report the observation of a remarkably strong microwave photoresistivity effect in a high-mobility two-dimensional electron system subject to a weak magnetic field and low temperature.
The effect manifests itself as a giant microwave-induced resistivity peak which, in contrast to microwave-induced resistance oscillations, appears {\em only} near the second harmonic of the cyclotron resonance and {\em only} at sufficiently high microwave frequencies.
Appearing in the regime linear in microwave intensity, the peak can be more than an order of magnitude stronger than the microwave-induced resistance oscillations and cannot be explained by existing theories.

\end{abstract}
\maketitle

Transport properties of high-mobility two-dimensional electron systems (2DESs) subject to a weak magnetic field $B$ and low temperature $T$ can be modified dramatically by microwave radiation,\citep{miro:exp1} thermally excited acoustic phonons,\citep{piro:1} dc electric fields,\citep{hiro:1} or their combinations.\citep{comb:1} 
In either case, the 2DES reveals a specific class of $1/B$-periodic resistance oscillations, which persist down to magnetic fields much lower than the onset of the Shubnikov-de Haas oscillations (SdHOs).

In irradiated 2DESs, such oscillations, usually called microwave-induced resistance oscillations (MIROs), are controlled by a dimensionless parameter, $\eac = \omega/\oc$, where $\omega=2\pi f$ is the microwave frequency and $\oc=e B/m^*$ is the cyclotron frequency of an electron with an effective mass $m^*$.
The resistivity can be expressed as $\rho_\omega = \rho+\delta\rho_\omega$, where $\rho$ is the resistivity of nonirradiated 2DES and $\delta\rho_\omega (\eac)$ is a sign-alternating photoresistivity.
According to the ``displacement'' model,  
$\delta\rho_\omega$ originates from the radiation-induced impurity-assisted transitions between the Landau levels.\citep{miro:th:disp}
In another mechanism, known as ``inelastic'', microwaves create a nonequilibrium distribution of electron states which, in turn, translates to the oscillatory $\delta\rho_\omega$.\citep{miro:th:in}
In the regime of overlapping Landau levels both models give
\be
\frac {\delta \rho_\omega}{\rho} \propto - {\po} \lambda^2\eac \sin (2\pi\eac).
\label{miro}
\ee
Here, $\po\propto \omega^{-4}$ is the dimensionless parameter proportional to the microwave power, $\lambda = \exp(-\pi/\oc\tauq) $ is the Dingle factor, and $\tauq$ is the quantum lifetime.
Even though Eq.\,(\ref{miro}) predicts the MIRO maxima at $\eac^{n+} \simeq n - 1/4$ ($n=1,2,3,\dots$), experimentally the lower order peaks are often found at $\eac^{n+}\simeq n - \varphi$, with $0<\varphi<1/4$.

In a very clean 2DES, MIRO minima can evolve into zero-resistance states which are believed to originate from the absolute negative resistance and its instability with respect to formation of current domains.\citep{zrs}
As a result, negative photoresistivity never exceeds the dark resistivity by absolute value.
In contrast, positive photoresistivity has no underlying instabilities and was routinely found to exceed the dark resistivity.
In addition to MIROs, a remarkably strong and narrow photoresistivity peak was recently observed in close proximity to the cyclotron resonance.\citep{smet:2005s}
This peak showed thresholdlike dependence on microwave power and was explained by the bolometric effect due to resonant heating of electrons.

In this paper, we report on another unusually strong microwave photoconductivity effect in a high-mobility 2DES. 
This effect manifests itself as a giant photoresistance peak, which emerges {\em only} near the {\em second} harmonic of the cyclotron resonance.
Similar to MIROs, the peak exhibits linear dependence on microwave intensity and quickly disappears with increasing temperature.
However, in contrast to MIROs, the amplitude of which quickly decays with increasing frequency (as $1/\omega^4$), the giant peak is observed {\em only} when the frequency is sufficiently high.
While the peak roughly coincides with the second MIRO maximum, it can be more than an order of magnitude {\em stronger} than MIROs. 
Understanding the nature of such a dramatic effect remains a subject of future studies.

While the effect was observed in several 2DESs, the data presented here were collected using a Hall bar (width $w=100$ $\mu$m) etched from a symmetrically doped GaAs/AlGaAs quantum well.
After a brief low-temperature illumination, the density and the mobility at $T=0.5$ K were $n_e\simeq 3.3\times 10^{11}$ cm$^{-2}$ and $\mu\simeq 1.1 \times 10^{7}$ cm$^{2}$/Vs, respectively.
Microwave radiation was generated by Gunn and backward wave oscillators. 
Measurements were done using a quasi-dc lock-in technique at bath temperatures $T$ from $0.5$ K to $4.0$ K.

\begin{figure}[t]
\includegraphics{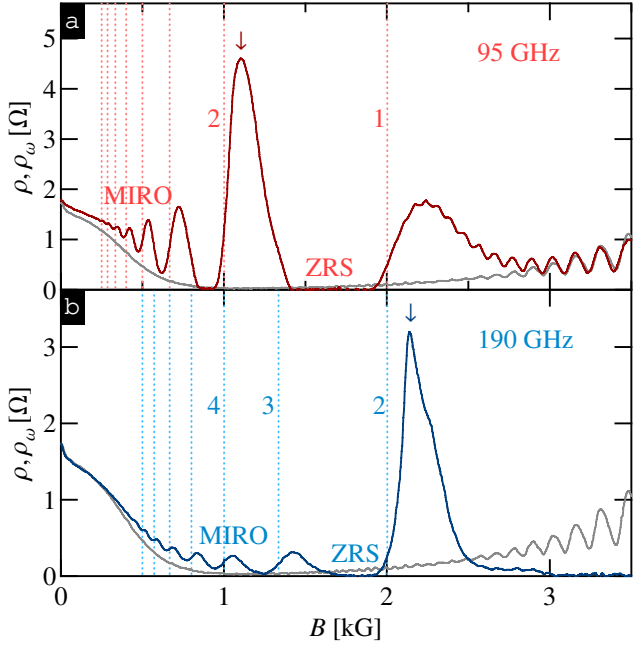}
\caption{(color online) (a), (b) Magnetoresistivities measured with ($\rho_\omega(B)$, dark curves) and without ($\rho(B)$, light curves) microwave irradiation of $f=95$ and 190 GHz  at $T \simeq 0.5$ K. 
Vertical lines are marked by integer $\eac$.
}
\vsp
\label{fig1}
\end{figure}
In Fig.\ref{fig1}\,(a) and 1\,(b) we show the magnetoresistivity $\rho_\omega(B)$ acquired at $T \simeq 0.5$ K under microwave irradiation (dark curves) of frequency $f=95$ and 190 GHz.
For comparison, each panel also includes the magnetoresistivity $\rho(B)$ obtained without microwave irradiation (light curves).
Dark resistivity $\rho(B)$ shows a strong negative magnetoresistance effect at low magnetic fields; at $B\simeq 1$ kG the dark resistivity is reduced by nearly two orders of magnitude compared to its value at $B=0$.
At higher $B$, magnetoresistance becomes positive and SdHOs appear.
Under microwave irradiation, the data show both MIROs and zero-resistance states.
However, our main focus is the so-called $\xt$ peak (cf.\,$\downarrow$) near $\eac=2$, which, at least in the case of $f=190$ GHz, is distinct from MIROs.

We first notice that direct examination of Fig.\,\ref{fig1} reveals that the $\xt$ peak is considerably stronger than other oscillations.
Second, the appearance of the peak is uniquely tied to the second harmonic of the cyclotron resonance as neither the third nor fourth harmonic shows similar features.
Finally, comparing the 190 GHz to the 95 GHz data we find that MIROs are suppressed considerably, mostly due to $\omega^{-4}$ decay of $\po$ [cf.\,\req{miro}].
On the other hand, the $\xt$ peak becomes even more pronounced, suggesting a frequency dependence, which is clearly inconsistent with $\omega^{-4}$.
This characteristic frequency dependence might explain why the $\xt$ peak was not detected in earlier studies employing lower frequencies.
Other necessary conditions for the observation of this unusual peak are sufficiently high mobility and low temperature.

\begin{figure}[t]
\includegraphics{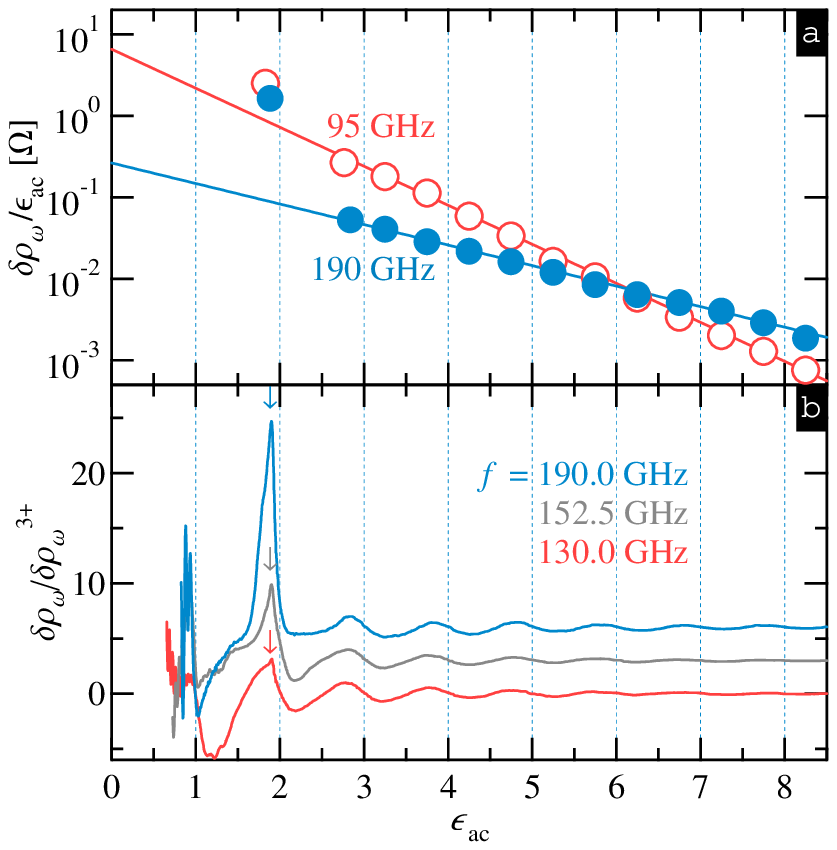}
\caption{(color online) (a) Normalized oscillation amplitude $\delta\rho_\omega/\eac$ versus $\eac$ at $f=95$ and 190 GHz. 
Solid lines show exponential decay $\delta\rho_\omega/\eac \propto \exp(-\eac/f\tauq)$ with $\tauq = 9.1$ ps.
Extrapolation of these lines to $\eac=0$ gives the ratio of microwave intensities, $\po(95\,{\rm GHz})/\po(190\,{\rm GHz}) \simeq 30$.
(b) Photoresistivity normalized to the $\eac^{3+}$ peak $\delta\rho_\omega/\delta\rho_\omega^{3+}$ versus $\eac$ for $f=130$, 152.5, and 190 GHz (bottom to top) at $T\simeq 0.5$ K.
The traces are vertically offset for clarity by 3.
}
\vsp
\label{dingle}
\end{figure}
To quantitatively compare the $\xt$ peak to MIROs, we present the normalized oscillation amplitude, $\delta\rho_\omega/\eac$, as a function of $\eac$ in Fig.\,\ref{dingle}\,(a).
For $\eac \geq 3$, both data sets exhibit anticipated exponential decay, $\delta\rho_\omega/\eac \propto \lambda^2 = \exp(-\eac/f\tauq)$, as illustrated by solid lines drawn with $\tauq = 9.1$ ps.
It is clear, however, that the magnitude at the $\xt$ peak significantly exceeds these dependences;
direct comparison shows that the photoresistance at the $\xt$ peak is enhanced by roughly a factor of 3\,(18) for $f=95\,(190)$ GHz.
Here, we should notice that, while for $\eac \gtrsim 3$, $\po$ can be treated as $\eac$ independent, it is expected to be enhanced considerably near the cyclotron resonance.\citep{khodas}
This enhancement can, in principle, increase the response near $\eac = 2$ by a factor of about 2.
If this correction is taken into account, the amplitude of the $\xt$ peak measured at $f=95$ GHz is in closer agreement with the MIRO amplitude but it is clearly not enough to explain the peak value at $f=190$ GHz.
We also notice that the $\xt$ peak at $f=190$ GHz is significantly sharper compared to a peak at $f=95$ GHz which has a shape similar to conventional MIROs.
We therefore concentrate on higher frequency data.

We now switch to the evolution of the $\xt$ peak with microwave frequency.
In Fig.\,\ref{dingle}\,(b) we present the photoresistivity for three frequencies (as marked), normalized to the $\eac^{3+}$ MIRO peak, $\delta\rho_\omega/\delta\rho_\omega^{3+}$ versus $\eac$.
Such normalization helps to account for the variation of the microwave intensity seen by our 2DES.
We observe that the $\xt$ peak (cf. $\downarrow$) grows with increasing $\omega$ confirming that its frequency dependence is inconsistent with that of MIROs.
Furthermore, as prescribed by Eq.\,(1), all the peaks, including the $\xt$ peak, are positioned near $\eac^{n+} = n - \phi$, with $\phi>0$.
This result is in contrast to \oncite{dai:2010}, which, based on absorption measurements, concludes that the $\xt$ peak occurs at $\eac = 2$.

\begin{figure}[t]
\includegraphics{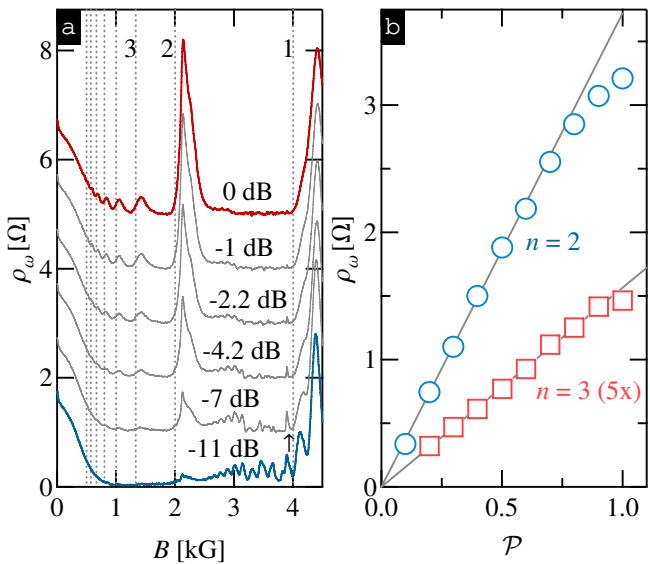}
\caption{(color online) (a) Magnetoresistivity $\rho_\omega (B)$ measured at $f=190$ GHz at different attenuations, from 0 dB to -11 dB. The traces are marked by attenuation factors and are vertically offset by 1 $\Omega$.
(b) Resistivity at the $\xt$ peak (circles) and $\eac^{3+}$ peak (squares, values multiplied by five) versus ${\mathcal P}$.
}
\vsp
\label{power}
\end{figure}
We next examine the power dependence of MIROs and the $\xt$ peak measured at $f=190$ GHz.
In Fig.\,\ref{power}\,(a) we present magnetoresistivity $\rho_\omega (B)$ acquired at selected attenuations, as marked, from 0 dB (top trace) to -11 dB (bottom trace). 
The attenuation steps were selected to roughly mimic a constant step in microwave intensity.
The traces are vertically offset by 1 $\Omega$ for clarity.
Figure\,\ref{power}\,(a) clearly shows that, with decreasing radiation intensity, MIROs gradually diminish and so does the $\xt$ peak.
We also note that, once the zero-resistance state disappears, the data reveal another rather weak but sharp peak just below the cyclotron resonance (cf.\,$\uparrow$).

To understand the observed evolution with microwave intensity, we present in Fig.\,\ref{power}\,(b) the resistance values at the peaks near $\eac=2$ (circles) and $\eac=3$ (squares, values multiplied by five) as a function of microwave intensity $\mathcal P$ (in units of intensity before attenuation). 
At not too high $\mathcal P$, both $\xt$ and $\eac^{3+}$ peaks show linear dependence on $\mathcal P$.
However, at highest $\mathcal P$, both dependencies show signs of saturation.
Such saturation may originate from the nonresonant heating of the 2DES by microwaves, which is manifested in the progressively stronger damping of the SdHOs with increasing $\mathcal P$.
Another possible origin for the sublinear dependence is the intrinsic nonlinearity of the photoresponse due to multiphoton processes.\citep{khodas}

\begin{figure}[t]
\includegraphics{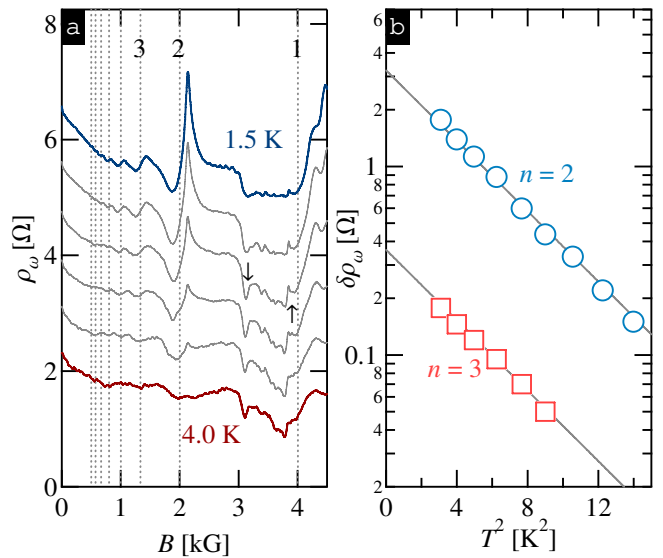}
\caption{(color online) (a) Magnetoresistivity $\rho_\omega (B)$ measured at $f=190$ GHz at ${\mathcal P} = 1$ and different temperatures, from 1.5 K (top) to 4.0 K (bottom), in step of 0.5 K. 
The traces are vertically offset by 1 $\Omega$.
(b) Photoresistivity $\delta \rho_\omega$ at the $\xt$ peak (circles) and $\eac^{3+}$ peak (squares) versus $T^2$.}
\vsp
\label{tempdep}
\end{figure}

We now turn to our results of a temperature-dependence study.
In Fig.\,\ref{tempdep}\,(a) we show the magnetoresistivity $\rho_\omega (B)$ measured at $f=190$ GHz and at different temperatures from $T=1.5$ K (top) to $T=4.0$ K (bottom), in steps of 0.5 K. 
For clarity, the traces are vertically offset by 1 $\Omega$.
With increasing $T$, both MIROs and the $\xt$ peak gradually weaken and eventually decay away.
Another interesting observation is a sharp photoresistivity minimum (cf.\,$\downarrow$), which emerges at the lower $B$ edge of the zero-resistance state developed between 3 and 4 kG at $T=1.5$ K.
Finally, similar to the low $\mathcal P$ data in Fig.\,\ref{power}\,(a) one observes a sharp peak at the higher $B$ edge of the zero-resistance state near $\eac \simeq 1$ (cf.\,$\uparrow$).

In Fig.\,\ref{tempdep}\,(b) we present the photoresistivity $\delta \rho_\omega$ at the $\xt$ peak (circles) and $\eac^{3+}$ peak (squares) versus $T^2$ and observe that both data sets are well described by $\delta\rho_\omega (T) \propto \exp(-T^2/T_0^2)$, with $T_0 \simeq 2.2$ K (cf.\,lines).
We note that similar behavior, recently observed for all classes of induced resistance oscillations, was attributed to electron-electron interactions.\citep{iro:temp} 
In the regime of separated Landau levels theory \citep{miro:th:in} predicts, up to a factor of the order of unity, $k_{B}T_{0} \simeq \sqrt{\Gamma \varepsilon_F/2\pi}$, where $k_B$ is the Boltzmann constant, $\varepsilon_F$ is the Fermi energy, and $2\Gamma$ is the Landau level width.
Using $\tauq \simeq 10^{-11}$ s and $2\Gamma = \hbar/\tauq$ yields $T_0 \simeq 3$ K, in agreement with experiment. 
We thus conclude that the $T$-dependence of the $\xt$ peak could also be explained by electron-electron interactions.

In summary, we reported on a novel microwave-induced resistivity peak emerging in a high-mobility 2DES at low temperatures.
Similar to MIROs, this peak grows linearly with power and decays exponentially with temperature, but is clearly of a different origin.
First, it appears {\em only} near the {\em second} harmonic of the cyclotron resonance and is not observed at other harmonics, regardless of the magnetic field.
Second, the peak appears {\em only} at sufficiently high microwave frequencies and its frequency dependence differs dramatically from MIROs, the amplitude of which decays as $\omega^{-4}$.
Finally, it can be more than an order of magnitude stronger than the microwave-induced resistance oscillations and more than two orders of magnitude larger than the dark resistivity.
This phenomenon cannot be explained by any of the existing theories and prompts for further developments in the field of nonequlibrium transport of quantum Hall systems.

We thank I. Dmitriev, R. Du, M. Dyakonov, M. Khodas, and B. Shklovskii for discussions and S. Hannas, G. Jones, J. Krzystek, T. Murphy, E. Palm, J. Park, D. Smirnov, and A. Ozarowski for technical assistance.
A portion of this work was performed at the National High Magnetic Field Laboratory, which is supported by NSF Cooperative Agreement No. DMR-0654118, by the State of Florida, and by the DOE.
The work at Minnesota was supported by the DOE Grant No. DE-SC0002567 (measurements at NHMFL) and by the NSF Grant No DMR-0548014 (low frequency measurements at Minnesota). 
The work at Princeton was partially funded by the Gordon and Betty Moore Foundation as well as the NSF MRSEC Program through the Princeton Center for Complex Materials (DMR-0819860).
A.T.H. acknowledges support by the University of Minnesota.

\end{document}